# Tongue Liminary Threshold Identification to Electrotactile Stimulation

Robineau Fabien[*]    Vuillerme Nicolas[*]    Jean-Pierre Orliaguet[*†]    Yohan Payan[*]

(*)TIMC IMAG Laboratory, CNRS-UMR 5525, France
(*†)LPNC Laboratory, CNRS-UMR 5105, France
E-mail: Fabien.Robineau@imag.fr

## Abstract

*Many applications use electrostimulation of the human skin to provide tactile sensation. The effect of electrotactile stimulations were studied on a 6x6 matrix of tactile electrodes placed on the anterior part of the tongue. The liminary threshold with continuous or discontinuous waveform and patterns with 2 or 4 electrodes was investigated.*

*The result suggest that for energy saving and to improve the yield, it would probably be better to use discontinuous stimulation with two electrode patterns.*

## 1. Introduction

Electrotactile stimulation can provide sensory information not available to damaged classical sensory system (vision, hearing, vestibular or somatosensory systems). Many electrotactile human machines interface (HMI) have been developed and applied to various surface areas [2]. The first development of electrotactile visual substitution systems (TVSS) was designed to provide distal spatial information to blind people via a 20x20 electrodes matrix [5]. Later Bach-y-Rita and his collaborators converged towards the electro-stimulation of the tongue surface, called "Tongue Display Unit" (TDU), which provides a practical HMI [4]. The human tongue is very sensitive, highly mobile and discriminative (spatial threshold < 2 mm). Moreover, the environment of the mouth offers a protected volume.

As concerns the tongue sensitivity, most authors agree that the tactile sensitivity threshold of the anterior part of the tongue, and particularly the tip, is lower than the other part [10, 9]. Moreover, it was showed that the mechano-receptive innervation is denser on the tip than in the other regions. However, few research has been investigated about the asymmetry of tongue sensitivity with respect to the median sulcus (right and left tongue sides) [13].

TIMC-IMAG laboratory recently developed medical electrotactile human machine interfaces using the TDU of Bach-y-Rita with a 6x6 matrix. These devices, which feasibility was evidenced, are used for disable persons assistance and for computer-aided surgery. Either the TDU provides electrotactile signal to supply punctual information, or the functioning is almost continuous to give orientation information. An application to prevent pressure ulcer formation in paraplegics, developed as an "alarm", provides via the TDU, tactile information about excess of pressure at the skin/seat interface [11]. In addition, a device to improve human ankle joint position sense was investigated [16]. On the other hand, electrotactile information comparable to continuous information is investigating. In this way, a biofeedback system aims at improving human balance control for people with visual or hearing impairments [15]. Also, is studied a system to provide to a surgeon, via the TDU device, orientation information to accuracy guide a needle until a target inside a body [14].

In these studies, a minimum of four electrode patterns were used because the subjects were able to discriminate without problem the stimuli. In parallel, a more ergonomic wireless 6x6 TDU device was recently developed in our laboratory. It is inserted in a dental retainer including microelectronics, antenna and power supply [14]. To increase the comfort of such an embedded device, the size and the energy consumption have to be even more reduced. The intensity on the matrix surface has to be as low as possible, but of course sufficient to be perceived by the subject (i.e. higher than the "liminary threshold" of the human tongue). Another option to decrease such intensity would be to reduce the number of activated electrodes. However, such a reduction should not decrease the discrimination of the electrotactile patterns.

The present study aims at evaluating the liminary thresholds of the tongue with continuous or discontinuous electrotactile stimulations among subjects, and comparing such thresholds with 2 or 4 electrode patterns.

## 2. Methods
### 2.1 Subjects

Two groups of ten subjects (age: 28,3 ± 3,9 years; body weight: 68,8 ± 10,2 kg; height: 174,6 ± 9,9 cm) voluntarily participated in this experiment. None of the subjects presented any history of sensory/motor or cognitive problem.

## 2.2 Apparatus

Electrotactile stimuli were delivered to the dorsum of the tongue via a ribbon TDU derived from the one developed by Bach-y-Rita [2, 4]. This electrotactile device consisted of an array of 36 tactile electrodes (6×6 matrix, radius: 0.7mm each), embedded in a 1,5×1,5 cm plastic strip (Fig. 1, left panel). The tip of the TDU was inserted in the oral cavity and held lightly between the lips, maintaining the array in close and permanent contact with the surface of the tongue (Fig. 1, right panel). A flexible cable, made of a thin (100 μm) strip of polyester material, connected the matrix to an external electronic device. This device delivered the electrical signals that activated the tactile receptors on the anterior superior part of the tongue. As Kaczmarek suggested [8], the frequency of the DC stimulating pulses was fixed to 50Hz for all trials and subjects. Because of the conductive properties of the saliva and the epidermis thickness, the TDU only required a 5-15V input voltage and a 0.4-4.0mA current. Each gold-plated circular electrode (Ø = 1.4mm, inter-centre distance = 2.3mm) received monophasic pulses.

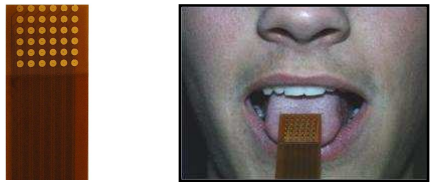

Fig. 1 On the left panel, the Tongue Display Unit. It constitutes of 2D electrode array (1,5 cm × 1,5 cm) including 36 gold-plated contacts each with 1.4 mm diameter, arranged in a 6×6matrix. On the right panel, the TDU placed on the tongue.

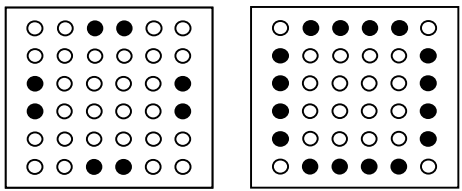

Fig. 2 Electrode Schemes of the four patterns couples tested on the anterior part of the tongue. Left panel: patterns with 2 electrodes; right panel: patterns with 4 electrodes.

Considering that each electrode could be independently activated, four tactile patterns couples were designed. As illustrated in Fig. 2, these patterns consist in the activation of either 2 or 4 electrodes located on the front, the left, the back and the right side of the matrix.

## 2.3 Task and procedure

Before the experiment, a set of instructions describing the different kind of patterns used in the study and the required task was orally presented to each participant. The TDU matrix was placed inside the mouth and in contact with the anterior superior surface of the tongue. The participants were asked to keep the TDU in the same position on her/his tongue as well as possible during the course of the experiment.

Two groups carried out the experiment. The first group performed the test with continuous waveform (CT condition) and the second group with discontinuous waveform (DCT condition). For each group, the patterns with two electrodes were first tested, followed by the patterns with four electrodes.

A psychophysical technique referred to as the "method of limits" [7] was employed in determining each pattern's liminary thresholds for each subject. An ascending threshold was established by beginning with a very low, previously tested stimulation threshold, gradually increasing by steps of 150 mV until the subject perceived a stimulation. A descending threshold was established by beginning with a stimulation clearly detectable (previously tested) and gradually decreasing by steps of 150 mV until the subject could no longer perceive any stimulation. A dedicated software administered the test of the liminary threshold detection. A total of 5 ascending and 5 descending series were alternatively run to each of the eight patterns for each subject. During the continuous stimulation test, the stimuli intensity was maintained 3s then increased or decreased as a function of the phase (ascending or descending). In the case of the discontinuous stimulation, the stimuli were sent during 1 s every 3 s in order to avoid a sensory adaptation and evoke a good recovery.

The subject was required to signal when she/he was absolutely certain of the appearance (ascending test) or the disappearance (descending test) of the lingual stimulation, by pressing a user-defined key on the laptop placed in front of her/him. To indicate the direction of the test, an up or down arrow was shown on the computer screen. A red point was also shown to specify when there was stimulation. Each testing session lasted about 2 hours.

## 2.4. Data analysis

The experiment yielded a total of 100 thresholds per pattern tested. The identified threshold for each pattern and for each subject was defined as the average of the median value for the 5 ascending responses and the median value for the 5 descending responses. The threshold values were calculated as the value at 50% of the 5 ascending or descending responses, with a linear interpolation supposing that the probability variation was linear between two values which surrounded 50% .

Data were computed and treated with Statistica statistics software. They were then submitted to four separated 2 Groups (Continuous vs Discontinuous) x 2 Electrode Patterns (2 electrodes vs 4 electrodes) analyses of variances (ANOVAs). The level of significance was set at 0.05.

## 3. Results

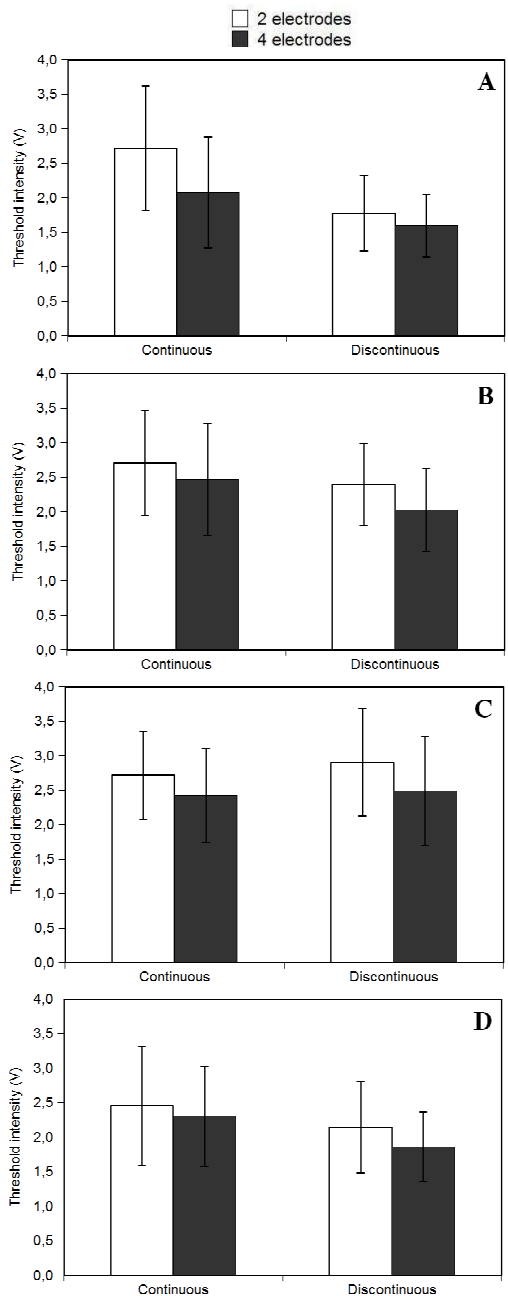

Fig. 3 Mean intensity liminary threshold for CT and DCT condition as a function of electrode number, 2 or 4, for A) the front, B) the left, C) the back, D) the right positions.

Figure 3 illustrates the mean intensity liminary threshold (± standard deviations) obtained for CT and DCT conditions as a function of number of activated electrodes (2 or 4), for the four front (A), left (B), back (C) and right (D) positions.

Results show a main effect for the group, for the front position (F (1;18)=15.01, p < 0.01, figure 3A). Liminary threshold is higher for the continuous than for the discontinuous condition.

Moreover, the results show a significant difference between 2 and 4 electrode patterns. For all positions, the patterns with 2 electrodes require an intensity threshold higher than the patterns with 4 electrodes (front, left, back and right, respectively F (1;18)=20.27, p < 0.001; F (1;18)=13.33, p < 0,01; F (1;18)=18.68, p < 0.001 and F (1;18)=16.13, p < 0.001).

Finally, the ANOVA shows a significant interaction between the groups and electrode numbers conditions for the front position (Fig. 3A), (F (1;18)=6.22, p < 0.03). Planned comparison analyses evidence a significant difference between two and four electrodes for continuous stimulation (F (1;18)=24.48, p < 0.001), whereas no significant difference was observed for the discontinuous stimulation (F (1;18)=2.01, p > 0.17).

## 4. Discussion

Tactile vibratory sensations are transmitted on the tongue surface via electrotactile stimulation. The electric current passing through the skin stimulates cutaneous afferent fibers at the location of the tactile electrodes. Our laboratory addressed various medical applications using the electrotactile feedback to supplement the damaged sensory information or provide accurate orientation information not easily accessible.

The front part of the tongue shows a significant difference for the continuous vs discontinuous conditions. As previously reported in the literature [9] [10] [12] [13], the sensitivity of the tongue tip is higher than all the others parts. Interestingly, our result further demonstrated that on the tip tongue, the liminary threshold is higher using continuous stimulations than discontinuous. This high tip tongue sensitivity of the tongue anterior part can explain that the accommodation effect is faster with the continuous than with discontinuous stimulations. Indeed, after accommodation, the induced sensation becomes less clear and less discriminable. Therefore, the performances of the subjects are reduced. An option to limit this adaptation could be to find the best stimulation frequency to improve the subjects performance (alarm or continuous guidance) and to optimize the tongue tactile neuronal adaptation/recovery in order to allow a regular use.

In this experiment, all the electrode number main effects were significant. For each position (front, left, back and right), the electrotactile liminary threshold was higher for two electrodes than for four electrodes. This result is quite natural and consistent with the literature [1]. However, it is interesting to compare the reduced intensity saved with respect to the electrode number activated to optimize the energy consumption. Although the two and four electrode patterns intensity liminary thresholds are very closed (only a 0,9 ratio between both), there are twice more activated electrodes between both patterns. Consequently, patterns with two electrodes should be preferred to patterns with four electrodes, assuming that the subjects are still able to efficiently perceive such patterns.

Moreover, the observation of a significant interaction between both conditions (CT/DCT vs Electrode number) on the front part shows that the liminary threshold difference between two and four electrode patterns observed for the continuous stimulation is not present under discontinuous stimulation. These results suggest that, for energy saving and to improve the yield, it would probably be better to use discontinuous stimulation with two electrode patterns.

Obviously, complementary studies should be performed to investigate more quantitatively lingual electrotactile stimulation. Psychophysical experiments have shown that it possible to provide electrotactile sensation on the tongue with a very light current. Also, further experiments could allow optimising the comfortability threshold of the wireless TDU for a daily use.

## Acknowledgement

This research was supported by the Gravit ("Grenoble Valorisation") consortium.

## References

[1] Ajdukovic, D., "The relationship between electrode area and sensory qualities in lelectrical tongue stimulation", Acta Otolaryngol, Stockholm, 1984.

[2] Bach-y-Rita, P., Tyler, M.E., and Kaczmarek, K.A., "Seeing with the brain," Int. J. Hum-Comput. Int., Vol. 15, pp. 285-295, 2003.

[3] Bach-y-Rita, P. and Kaczmarek, K., "Tongue Placed Tactile Output Device," U.S. Patent 6,430,450 B1, 2002.

[4] Bach-y-Rita, P., Kaczmarek, K.A., Tyler, M.E., and Garcia-Lara, J., "Form perception with a 49-point electrotactile stimulus array on the tongue," J. Rehabil. Res. Dev., Vol. 35, pp. 427-430, 1998.

[5] Bach-y-Rita, P., Collins, C.C., Saunders, and al., "Vision substitution by tactile image projection", Nature, Vol. 221, pp. 963-964, 1969.

[6] Collins, C. and Bach-y-Rita, P., "Transmission of pictorial information through the skin," Advances in Biology and Med. Phys., Vol. 14, pp. 285-315, 1973.

[7] Gescheider, G.A., "Psychophysics: The Fundamentals", Lawrence Erlbaum Associates, Mahwah, NJ, 1997.

[8] Kaczmarek, K.A., "Electrotactile adaptation on the abdomen: Preliminary results," IEEE Trans. Rehab. Eng., Vol.8, pp. 499-505, 2000.

[9] Maeyaman T., Plattign K.H., "Minimal two-point discrimination in human tongue and palate", Am J Otolaryngol, Vol. 10, pp. 342-4, 1989.

[10] Marlow, C.D., Winkemann, R. K., Gibilsco, J.A. General sensory innervation of the human tongue. Anat. Rec., Vol. 152:, pp. 503-512, 1965.

[11] Moreau-Gaudry, A, Prince, A., Demongeo,t J. & Payan Y., "A New Health Strategy to Prevent Pressure Ulcer Formation in Paraplegics using Computer and Sensory Substitution via the Tongue", Studies in Health Technology and Informatics, Vol. 124, pp. 926-931, 2006.

[12] Pleasonton, A.K., "Sensitivity of the tongue to electrical stimulation", J Speech Hear Res., Vol. 13(3), pp. 635-44, 1970.

[13] Ringel, R. L. and Ewanowski, S.J., "Oral perception. I. Two-point discrimination", J. Speech Hear. Res., Vol. 8, pp. 389-398, 1965.

[14] Robineau, F., Boy, F., Orliaguet, J-P., Demongeot, J. & Payan, Y. "Guiding the surgical gesture using an electro-tactile stimulus array on the tongue: A feasibility study", IEEE Transactions on Biomedical Engineering, Vol. 54(4), pp. 711-717, 2007.

[15] Vuillerme, N, Chenu, O, Demongeot, J. & Payan, Y. "Controlling posture using a plantar pressure-based, tongue-placed tactile biofeedback system", Experimental Brain Research, Vol. 179, pp. 409-414, 2007.

[16] Vuillerme N., Chenu O., Demongeot J. & Payan Y., "Improving human ankle joint position sense using an artificial tongue-placed tactile biofeedback", Neuroscience Letters, Vol. 405, pp. 19-23, 2006.